\documentclass{amsart}
\usepackage{amsmath}
\usepackage{amsfonts}
\usepackage{amsthm, upref}
\usepackage{graphicx}
\usepackage{enumerate}
\usepackage[usenames, dvipsnames]{color}
\usepackage{url}
\usepackage{tikz}
\usepackage{array}

\usetikzlibrary{calc, intersections}
\usetikzlibrary{decorations.markings}
\usetikzlibrary{decorations.pathreplacing}
\input xy
\xyoption{all}

\newcommand{\comment}[1]{}
\newcommand{\eq}{\begin{equation}}
\newcommand{\en}{\end{equation}}

\begin{document}

\theoremstyle{plain}
\newtheorem{theorem}{Theorem}[section]
\newtheorem{lemma}[theorem]{Lemma}
\newtheorem{proposition}[theorem]{Proposition}
\newtheorem{corollary}[theorem]{Corollary}

\theoremstyle{definition}
\newtheorem{definition}[theorem]{Definition}
\newtheorem{asmp}[theorem]{Assumption}
\newtheorem{notn}[theorem]{Notation}
\newtheorem{problem}[theorem]{Problem}

\theoremstyle{remark}
\newtheorem{remark}[theorem]{Remark}
\newtheorem{example}[theorem]{Example}
\newtheorem{clm}[theorem]{Claim}
\newtheorem{assumption}[theorem]{Assumption}

\numberwithin{equation}{section}

\title[Energy, entropy, and arbitrage]{Energy, entropy, and arbitrage}

\author{Soumik Pal}
\address{Department of Mathematics\\ University of Washington\\ Seattle, WA 98195}
\email{soumikpal@gmail.com}

\author{Ting-Kam Leonard Wong}
\address{Department of Mathematics\\ University of Washington\\ Seattle, WA 98195}
\email{tkleonardwong@gmail.com}


\date{\today}
\begin{abstract}
We introduce a pathwise approach to analyze the relative performance of an equity portfolio with respect to a benchmark market portfolio. In this energy-entropy framework, the relative performance is decomposed into three components: a volatility term, a relative entropy term measuring the distance between the portfolio weights and the market capital distribution, and another entropy term that can be controlled by the investor by adopting a suitable rebalancing strategy. This framework leads to a class of portfolio strategies that allows one to outperform, in the long run, a market that is diverse and sufficiently volatile in the sense of stochastic portfolio theory. The framework is illustrated with several empirical examples.
\end{abstract}

\maketitle

\section{Introduction} \label{sec:intro}
\subsection{Rebalancing and volatility pumping}
\begin{figure}[t]
\begin{tikzpicture}[scale = 0.5]
\draw [->] (0, 0) -- (7, 0);
\draw [->] (0, 0) -- (0, 5);
\node [right] at (7, 0) {time};
\node [above] at (0, 5) {price};
\node [left] at (0, 4) {$\$1$};
\node [left] at (0, 2) {$\$0.5$};

\draw[blue] (0, 4) -- (6, 4);
\draw[red] (0, 4) -- (1, 2);
\draw[red] (1, 2) -- (2, 4);
\draw[red] (2, 4) -- (3, 2);
\draw[red] (3, 2) -- (4, 4);
\draw[red] (4, 4) -- (5, 2);
\draw[red] (5, 2) -- (6, 4);
\node[blue, above, thick] at (3, 4) {asset 2};
\node[red, below, thick] at (3, 2) {asset 1};
\node[below] at (0, 0) {$0$};
\node[below] at (2, 0) {$2$};
\node[below] at (4, 0) {$4$};
\node[below] at (6, 0) {$6$};

\draw [->] (11, 0) -- (18, 0);
\draw [->] (11, 0) -- (11, 5);
\node [right] at (18, 0) {time};
\node [above] at (11, 5) {price};
\node [left] at (11, 4) {$\$1$};
\node [left] at (11, 2) {$\$0.5$};
\node [left] at (11, 1) {\small $\$0.25$};
\node [left] at (11, 0.5) {\tiny $\$0.125$};

\draw[blue] (11, 4) -- (17, 4);
\draw[red] (11, 4) -- (12, 2);
\draw[red] (12, 2) -- (13, 1);
\draw[red] (13, 1) -- (14, 0.5);
\draw[red] (14, 0.5) -- (15, 1);
\draw[red] (15, 1) -- (16, 2);
\draw[red] (16, 2) -- (17, 4);
\node[blue, above, thick] at (14, 4) {asset 2};
\node[red, below, thick] at (14, 2) {asset 1};
\node[below] at (11, 0) {$0$};
\node[below] at (13, 0) {$2$};
\node[below] at (15, 0) {$4$};
\node[below] at (17, 0) {$6$};

\end{tikzpicture}
\vspace{-0.2cm}
\caption{Illustration of volatility pumping. Asset 2 is cash and asset 1 either goes up by a factor of $2$ or goes down by a factor of $0.5$. Six periods are shown in the figure. Left: The price pattern is $-+-+-+$. Right: The price pattern is $---+++$.} \label{fig:example1}
\end{figure}
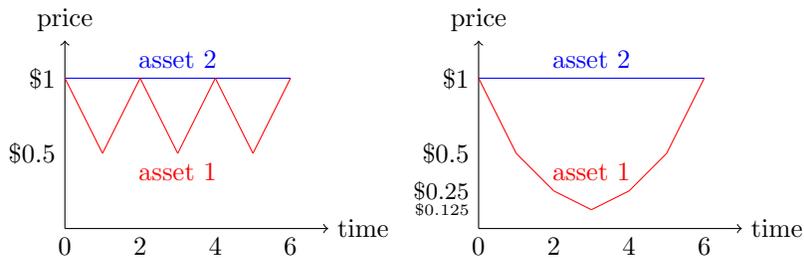

Rebalancing and volatility pumping are popular topics in mathematical finance and investment management. To recall a classic example (see for example \cite[Example 15.2]{L98} and \cite{D07}), consider two assets whose prices fluctuate as follows (see Figure \ref{fig:example1} (left)). Asset $1$ earns $-50\%$ return for all odd periods and $100\%$ return for all even periods. On the other hand, Asset $2$ is a risk-free asset whose return is always $0\%$.

If one buys and holds any of the two assets, clearly no long term growth will be created. Nevertheless, if the investor {\it rebalances} the portfolio so that equal amount of capital is invested in the two assets at the beginning of each period, the resulting {\it equal-weighted portfolio} outperforms any buy-and-hold portfolio exponentially in time. To see why, note that the return of the equal-weighted portfolio in the first period is
\[
\frac{1}{2} \times (-50\%) + \frac{1}{2} \times 0\% = -25\%.
\]
and the return for the second period is
\[
\frac{1}{2} \times 100\% + \frac{1}{2} \times 0\% = 50\%,
\]
Over two periods the growth of the portfolio is the product of the `down factor' $0.75$ and the `up factor' $1.5$:
\[
0.75 \times 1.5 = 1.125.
\]
This is strictly larger than $1$, and compounding gives exponential outperformance.

This example illustrates that systematic rebalancing is capable of capturing profit `from volatility' even when the underlying assets experience zero growth. Intuitively, this is because rebalancing has a built-in {\it buy-low-sell-high effect} which is absent in buy-and-hold portfolios. In practice, it has been observed (see \cite{PPA12, PUV12}) that a rebalancing portfolio often outperforms a capitalization-weighted benchmark portfolio.

In spite of the large volume of literature on the theory and practice of rebalancing and volatility pumping (see \cite{FS82, FM07, DESK08, PR11, MTZ11, CZ14, H14} and the references therein), there is some confusion among academics and practitioners. This, we believe, is due to the fact in most theoretical analyses of rebalancing very strong assumptions are imposed on the dynamics of asset prices, which give a false impression that rebalancing only works in those situations. An important case in point is the (wrong) assertion that rebalancing is profitable only when the underlying price changes are negatively correlated like in the above example \cite{CZ14}. There is thus a strong need to study the precise conditions under which rebalancing or volatility pumping beats a capitalization-weighted portfolio.



\subsection{Stochastic portfolio theory}
Important progress towards this question has been made by Stochastic Portfolio Theory (SPT) (see \cite{F02} and \cite{FKSurvey} for an introduction) which is a descriptive theory of equity market and portfolio selection. To fix ideas, consider an equity market consisting of $n$ stocks. If $X_i(t)$ is the market capitalization of stock $i$ at time $t$, the ratio
\begin{equation} \label{eqn:marketweight}
\mu_i(t) = \frac{X_i(t)}{X_1(t) + \cdots + X_n(t)}
\end{equation}
is called the {\it market weight} of stock $i$ and represents the proportion of total market capital in stock $i$. The vector $\mu(t) = \left(\mu_1(t), \ldots, \mu_n(t)\right)$ of market weights takes values in the open unit simplex
\[
\Delta_n = \left\{p = \left(p_1, \ldots, p_n\right): p_i > 0, \sum_{i = 1}^n p_i = 1\right\}.
\]
If at each period one invests according to the market weights, the resulting portfolio is called the {\it market portfolio}. The market portfolio is a capitalization-weighted portfolio whose value represents the overall performance of the market. Accordingly, it is frequently taken as an investment benchmark. For example, the S\&P500 index may be taken as a proxy of the market portfolio in the US stock market, and the MSCI EAFE index is a benchmark of the global equity (ex-US and Canada) market.

Rearranging the market weights from the largest to the smallest, we get the {\it capital distribution} of the market: $\mu_{(1)}(t) \geq \cdots \geq \mu_{(n)}(t)$. Stochastic portfolio theory exploits the fact that the distribution of capital exhibits remarkable stability over long periods (see \cite[Chapter 4]{F02}). A particular property, called {\it diversity}, states that the biggest company cannot dominate the market, i.e., $\mu_{(1)}(t) \leq 1 - \delta$ for all $t$, for some $\delta > 0$ (see \cite[Chapter 2]{F02}). Making use of this stability, the theory identifies a subset of a family of portfolios -- {\it functionally generated portfolios} -- which is capable of profiting from volatility by systematic rebalancing and outperforming the market portfolio over sufficiently long horizons. These portfolios are called {\it relative arbitrages} with respect to the market under the conditions of {\it diversity} and {\it sufficient volatility}. Each functionally generated portfolio is given by a deterministic function which assigns a portfolio vector $\pi(t)$ to the current market weight vector $\mu(t)$. An example is the {\it diversity-weighted portfolio} whose portfolio weights are given by
\begin{equation} \label{eqn:diversity}
\pi_i(t) = \frac{\mu_i^{\lambda}(t)}{\sum_{j = 1}^n \mu_j^{\lambda}(t)}, \quad i = 1, \ldots, n,
\end{equation}
where $0 < \lambda < 1$ is a fixed parameter. See \cite{F98} for its practical applications.
 
These remarkable results (see \cite[Chapter 2]{FKSurvey} for precise statements) were first obtained by Fernholz in a continuous time setting where stock prices are modeled by continuous It\^{o} processes. It is natural to ask whether the same results hold in discrete-time (after all portfolios cannot be rebalanced continuously in practice) and whether there are portfolios, not functionally generated, that beat the market under similar assumptions on diversity and sufficient volatility \cite[Remark 11.5]{FKSurvey}. Since the original proofs are based on stochastic calculus, an alternative approach is required.

\subsection{Our contributions}
The purpose of this paper is to introduce an information-theoretic, pathwise and model-free framework for analyzing the performance of any portfolio relative to the market portfolio, with an emphasis on profiting from market volatility. We call this the {\it energy-entropy framework}. This paper is the first of a series of papers \cite{PW14, W14, BVW15, W15} devoted to a pathwise approach to relative arbitrage, functionally generated portfolio and stochastic portfolio theory. In particular, \cite{BVW15} gives a more accessible treatment of the energy-entropy framework for practitioners and contains more empirical examples.

We summarize our contributions as follows. First, we provide a treatment of the basic results of stochastic portfolio theory in discrete time, showing that no stochastic modeling assumptions are required. Instead of analyzing specific statistical models of stock prices, the basic unit of our analysis is individual paths of stock prices. The absence of probability considerations allows us to focus on the path properties of stock prices that are relevant to rebalancing. Second, we derive a pathwise decomposition formula for the performance of \textit{any} portfolio relative to the market portfolio; it quantifies the profit or loss of the portfolio due to rebalancing and market movement. While decomposition formulas for functionally generated portfolios (see \cite[Theorem 3.1.5]{F02}) in continuous time stochastic portfolio theory are also pathwise, their derivation depends on stochastic calculus. Here, our derivations are completely free of probabilistic considerations. Third, this framework leads to a class of portfolio strategies that are capable of beating a diverse and sufficiently volatile market in the long run; they are not functionally generated (thus answering \cite[Remark 11.5]{FKSurvey}) and are more flexible. We call them {\it energy-entropy portfolios}. Last but not least, this framework is also suitable for analyzing hierarchical portfolios (i.e., fund of funds, see Section \ref{sec:hierarchical}). The energy-entropy framework is implemented in the \verb"R" package \verb"RelValAnalysis" (Relative Value Analysis) available on \verb"CRAN".

\medskip

The main idea of the mechanism underlying rebalancing and volatility pumping can be understood from a simple modification of the first example (it also shows that negative correlation is not required for rebalancing to work). Suppose that the returns of asset 1 are resuffled over time as in Figure \ref{fig:example1} (right). Now except the fourth period the price changes are positively correlated. The growth of the equal-weighted portfolio over the six periods remains unchanged because we can {\it rearrange} the factors:
\[
0.75 \times 0.75 \times 0.75 \times 1.5 \times 1.5 \times 1.5 = (0.75 \times 1.5)^3.
\]
Thus, the key driver of the long term growth of the rebalancing portfolio is the number of times the growth factor $0.75 \times 1.5$ can be matched. As it turns out, this is closely related to the concept of sufficient volatility. Our framework generalizes this idea for any number of assets and any dynamic portfolio strategy.

\subsection{Pairs trading} The special case of two assets often goes by the name of pairs trading. Widely used in practice \cite{GGR}, it is often the first example of statistical arbitrage that comes to mind. Unfortunately, although pairs trading has been in practice for over 30 years, its analysis often make use stringent assumptions such as mean reversion of the relative price of one asset with respect to the other. As we show by a binary tree model in Section \ref{sec:binary}, pairs trading is a special case of the general stochastic portfolio theory and its efficacy is determined just by diversity and relative volatility of one asset with respect to the other.  The advantage of our approach is that the above claim is shown pathwise for every discrete-time path without any assumptions or statistical modeling regarding its future behavior.

\subsection{Outline}
The rest of the paper is organized as follows. Section \ref{sec:prelim} describes the mathematical set up including the relative value of a portfolio strategy with respect to the market portfolio. In particular, Section \ref{sec:binary} gives a formal treatment of the two-asset example. It provides fresh insights into rebalancing and pairs trading and introduces connections with Fernholz's functionally generated portfolio. In Section \ref{sec:EE} we derive the main decomposition formulas of the energy-entropy framework. Several applications of the framework will be given in Section \ref{sec:applications}. In particular, we introduce a class of portfolio strategies called {\it energy-entropy portfolios}. 
In Section \ref{sec:empirical} we illustrate the performance of these portfolios using actual data.
Finally, in Section \ref{sec:hierarchical}, we extend our analysis to a portfolio of portfolios. We show our information-theoretic framework neatly attributes gains from rebalancing to the various levels of the hierarchy (i.e., rebalancing within portfolios vs.~rebalancing among portfolios) in the spirit of ANOVA decomposition in classical statistics.

\section{Preliminaries} \label{sec:prelim}
\subsection{Market capital distribution}
We consider an equity market in discrete time consisting of $n \geq 2$ stocks. Although we restrict ourselves to equity markets, other asset classes can be used for the following analysis as long as we interpret the market portfolio properly. For instance, we may interpret each `stock' as an industrial sector or a country equity index.

Let $X_i(t) > 0$ be the capitalization of stock $i$ at time $t$. We assume that each stock has a single outstanding share, so $X_i(t)$ may also be regarded as the price of stock $i$. The {\it market weight} of stock $i$ at time $t$ is defined by \eqref{eqn:marketweight}. In our idealized model all capital changes are driven by stock returns. Thus, if $R_i(t)$ represents the return of stock $i$ over the time interval $[t, t + 1]$, the market weights can be updated by the formula
\begin{equation} \label{eqn:muupdate}
\begin{split}
\mu_i(t + 1) &= \frac{X_i(t)(1 + R_i(t))}{X_1(t)(1 + R_1(t)) + \cdots + X_n(t)(1 + R_n(t))} \\
  &= \frac{\mu_i(t)(1 + R_i(t))}{\mu_1(t)(1 + R_1(t)) + \cdots + \mu_n(t)(1 + R_n(t))}.
\end{split}
\end{equation}
We visualize the evolution of the market as a sequence $\{\mu(t)\}_{t = 0}^{\infty}$ of market weight vectors with values in $\Delta_n$. In our pathwise approach, we do not assume that the sequence $\{\mu(t)\}_{t = 0}^{\infty}$ is a realization of a stochastic process. We take the path as it is and investigate what path properties are required for the existence of relative arbitrage opportunities.

\subsection{Relative value of portfolios}
A {\it portfolio weight vector} is an element of the closed unit simplex $\overline{\Delta}_n$ which is the closure of $\Delta_n$ in ${\mathbb{R}}^n$. The components of $\pi$ represent the proportions of capital invested in the available assets. In particular, all portfolios considered are all-long and fully invested in the equity market. A {\it portfolio strategy} is a sequence $\pi = \{\pi(t)\}_{t = 0}^{\infty}$ of portfolio weight vectors chosen sequentially based on historical prices and other information of the investor. Given $\pi$, consider the corresponding self-financing portfolio with initial investment $\$1$. If $Z_{\pi}(t)$ denotes the value of this portfolio, we have
\begin{equation} \label{eqn:portfoliovalue}
Z_{\pi}(t + 1) = Z_{\pi}(t) \left(1 + \sum_{i = 1}^n \pi_i(t) R_i(t)\right),
\end{equation}
where $R_i(t)$ is the return of stock $i$ over the interval $[t, t + 1]$. If $\pi_i(t) \equiv \mu_i(t)$, the resulting portfolio is called the {\it market portfolio} and will be denoted by $\mu$. It is easy to check that
\[
Z_{\mu}(t) = \frac{X_1(t) + \cdots + X_n(t)}{X_1(0) + \cdots + X_n(0)}.
\]

The objective of this paper is to analyze the performance of a portfolio relative to the market portfolio. This will be measured by the {\it relative value}. For simplicity transaction costs are not included.

\begin{definition} [Relative value]
Given a portfolio strategy $\pi$, its relative value (with respect to the market portfolio) is the ratio
\[
V_{\pi}(t) = \frac{Z_{\pi}(t)}{Z_{\mu}(t)}.
\]
\end{definition}

The following lemma shows that the evolution of the relative value depends only on the market weights.

\begin{lemma} \label{lem:relativevalue}
The relative value of any portfolio strategy $\pi$ satisfies $V_{\pi}(0) = 0$ and
\begin{equation} \label{eqn:relativevalue}
\frac{V_{\pi}(t + 1)}{V_{\pi}(t)} = \sum_{i = 1}^n \pi_i(t) \frac{\mu_i(t + 1)}{\mu_i(t)}.
\end{equation}
\end{lemma}
\begin{proof}
Since $Z_{\pi}(0) = Z_{\mu}(0) = 1$, it is clear that $V_{\pi}(0) = 1$. We have
\[
\frac{Z_{\pi}(t + 1)}{Z_{\pi}(t)} = \sum_{i = 1}^n \pi_i(t) \left(1 + R_i(t)\right)
\]
and
\[
\frac{Z_{\mu}(t + 1)}{Z_{\mu}(t)} = \sum_{i = 1}^n \mu_i(t) \left(1 + R_i(t)\right).
\]
By \eqref{eqn:muupdate}, we have
\begin{equation}
\begin{split}
\frac{V_{\pi}(t + 1)}{V_{\pi}(t)} &= \frac{Z_{\pi}(t + 1) / Z_{\pi}(t)}{Z_{\mu}(t + 1) / Z_{\mu}(t)} \\
  &= \sum_{i = 1}^n \pi_i(t) \frac{1 + R_i(t)}{\sum_{j = 1}^n \mu_j(t) (1 + R_j(t))} \\
  &= \sum_{i = 1}^n \pi_i(t) \frac{\mu_i(t + 1)}{\mu_i(t)}.
\end{split}
\end{equation}
\end{proof}

A portfolio strategy $\pi$ is said to be {\it constant-weighted} if the portfolio vector $\pi(t)$ is constant over time. Abusing notations, a constant-weighted portfolio can be represented by a vector $\pi \in \overline{\Delta}_n$. 

\begin{remark}[Rebalancing] \label{re:rebalancing}
Most portfolio strategies, including the constant-weighted portfolios (where $\pi$ has at least two strictly positive components), require trading in order to maintain the desired portfolio weights. More precisely, suppose at time $t$ the portfolio vector is $\pi(t)$. By the consideration leading to \eqref{eqn:muupdate}, just before trading happens at time $t + 1$, the proportion of capital in stock $i$ is
\begin{equation} \label{eqn:implied}
\widetilde{\pi}_i(t + 1) = \frac{\pi_i(t) (1 + R_i(t))}{\sum_{j = 1}^n \pi_j(t) (1 + R_j(t))}.
\end{equation}
The {\it implied weights} $\widetilde{\pi}_i(t + 1)$ are sometimes called the {\it drifted weights} by portfolio managers, and properly speaking rebalancing is the trading which moves the portfolio weights from $\widetilde{\pi}(t + 1)$ to the new weights $\pi(t + 1)$ (instead of moving from $\pi(t)$ to $\pi(t + 1)$). With this terminology, a {\it buy-and-hold portfolio} -- such as the market portfolio $\mu$ -- is a portfolio strategy $\pi$ satisfying $\widetilde{\pi}(t + 1) = \pi(t + 1)$ for all $t$. 
\end{remark}

\subsection{Two-asset case} \label{sec:binary}
In this subsection we expand upon the two-asset example in the Introduction using a binomial tree model. The intuition gained will be useful in the general case. Suppose there are two assets whose prices are $X_1(t)$ and $X_2(t)$. Let
\[
Y(t) := \log \frac{X_1(t)}{X_2(t)}
\]
be a measure of relative price. We assume that $\Delta Y(t) = Y(t + 1) - Y(t)$ only takes the values $\sigma$ and $-\sigma$ where $\sigma > 0$ is a fixed constant. We think of $\sigma^2$ as the instantaneous volatility of the relative prices.

Given a portfolio strategy $\pi(t) = \left(\pi_1(t), \pi_2(t)\right)$, we let
\[
W_{\pi}(t) = \frac{Z_{\pi}(t)}{X_2(t) / X_2(0)}
\]
be the value of the portfolio $\pi$ relative to asset 2. So asset 2 serves the purpose of num\'{e}raire (in the example in the Introduction asset 2 is cash). Then $W_{\pi}(0) = 1$ and, by an argument similar to that in Lemma \ref{lem:relativevalue}, we have $W_{\pi}(0) = 1$ and
\[
\frac{W_{\pi}(t + 1)}{W_{\pi}(t)} = 1 + \pi_1(t) \left( e^{\Delta Y(t)} - 1 \right) =
\begin{cases}
1 + \pi_1(t) \left(e^{\sigma} - 1\right) &\mbox{if } \Delta Y(t) = \sigma, \\ 
1 + \pi_1(t) \left(e^{-\sigma} - 1\right) & \mbox{if } \Delta Y(t) = -\sigma.
\end{cases}
\]
These are the `up' and `down' factors (in Figure \ref{fig:example1} $\sigma = \log 2$). For any time $t$, we consider $W_{\pi}(t)$ as a product of these up and down factors:
\begin{equation} \label{eqn:productfactor}
W_{\pi}(t) = \prod_{s = 0}^{t - 1} \left(  1 + \pi_1(s) \left( e^{\Delta Y(s)} - 1 \right) \right).
\end{equation}

\begin{example}[Constant-weighted portfolio]
Let $\pi = (q, 1 - q)$ be a constant-weighted portfolio where $0 < q < 1$. For each pair of up and down moves, we get the contribution
\begin{equation} \label{eqn:kappa}
\kappa := \left(1 + q \left(e^{\sigma} - 1\right)\right)\left(1 + q \left(e^{-\sigma} - 1\right)\right) = 1 + q(1 - q) \left(e^{\sigma/2} - e^{-\sigma/2}\right)^2 > 1.
\end{equation}
Thus each matching creates a `growth factor' $\kappa$. Note that this quantity is maximized when $q = \frac{1}{2}$, i.e., the portfolio is equal-weighted.

Suppose there are $N(t)$ matchings up to time $t$ (see Figure \ref{fig:matching}). From Figure \ref{fig:matching}, it is clear that the number of unmatched moves is precisely $M(t) := |Y(t) - Y(0)| / \sigma$. In the product representation \eqref{eqn:productfactor}, for each pair of up and down moves we get the factor $\kappa$ given by \eqref{eqn:kappa}. Thus we get the decomposition
\begin{equation} \label{eqn:binarydecomposition}
\log W_{\pi}(t) = N(t) \kappa + M(t) \eta,
\end{equation}
where $\eta := \log \left(1 + q \left(e^{\pm \sigma} - 1\right)\right)$ depending on the sign of $Y(t) - Y(0)$. The decomposition \eqref{eqn:binarydecomposition} makes it clear that the constant-weighted portfolio outperforms asset 2 {\it if and only if} the number of matched moves is large relative to that of the unmatched moves. In particular, if $N(t) \uparrow \infty$ and $M(t) = o(N(t))$, then $W_{\pi}(t) \uparrow \infty$ as $t \uparrow \infty$.

\begin{figure}[t!]
\centering
\begin{tikzpicture}[scale = 0.4]

\draw [<->] (0, 6) -- (0, 0) -- (15, 0);
\draw       (0, 0) -- (0, -1.5);
\node [right] at (15, 0) {$t$};
\node [above] at (0, 6) {$Y(t)$};
\node [left] at (0, 1) {$Y(0)$};

\draw (0, 1) -- (1, 2); \draw [fill] (0, 1) circle [radius=0.1];
\draw (1, 2) -- (2, 3); \draw [fill] (1, 2) circle [radius=0.1];
\draw (2, 3) -- (3, 2); \draw [fill] (2, 3) circle [radius=0.1];
\draw (3, 2) -- (4, 3); \draw [fill] (3, 2) circle [radius=0.1];
\draw (4, 3) -- (5, 4); \draw [fill] (4, 3) circle [radius=0.1];
\draw (5, 4) -- (6, 3); \draw [fill] (5, 4) circle [radius=0.1];
\draw (6, 3) -- (7, 2); \draw [fill] (6, 3) circle [radius=0.1];
\draw (7, 2) -- (8, 1); \draw [fill] (7, 2) circle [radius=0.1];
\draw (8, 1) -- (9, 2); \draw [fill] (8, 1) circle [radius=0.1];
\draw (9, 2) -- (10, 1); \draw [fill] (9, 2) circle [radius=0.1];
\draw (10, 1) -- (11, 0); \draw [fill] (10, 1) circle [radius=0.1];
\draw (11, 0) -- (12, -1); \draw [fill] (11, 0) circle
[radius=0.1];
\draw (12, -1) -- (13, 0); \draw [fill] (12, -1) circle
[radius=0.1];
\draw (13, 0) -- (14, -1); \draw [fill] (13, 0) circle
[radius=0.1];
                           \draw [fill] (14, -1) circle
                           [radius=0.1];

\node [above] at (5, 5) {matched};
\draw[-latex,thick,black](4.8,5) to[out=240,in=120] (4.5,3.5);
\draw[-latex,thick,black](5.2,5) to[out=300,in=60] (5.5,3.5);

\node [below] at (5, 1) {matched};
\draw[-latex,thick,black](4.8,1) to[out=120,in=-60] (3.5,2.5);
\draw[-latex,thick,black](5.2,1) to[out=60,in=240] (6.5,2.5);

\node [above] at (12.5, 1.5) {unmatched};
\draw[-latex,thick,black](12.3,1.5) to[out=210,in=30] (10.5,0.5);
\draw[-latex,thick,black](12.7,1.5) to[out=310,in=90] (13.6,-0.5);
\end{tikzpicture}
\caption{Matching up and down moves in the binary tree model.}
\label{fig:matching}
\end{figure}
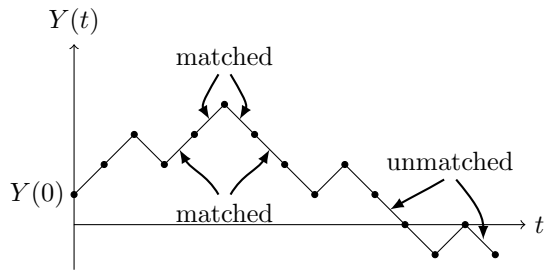
\end{example}

In the Appendix we consider rebalancing using state-dependent portfolio functions and relate them with Fernholz's functionally generated portfolio.

\section{Energy-entropy rebalancing} \label{sec:EE}
\subsection{Discrete excess growth rate} \label{sec:excessgrowth}
We begin our treatment of the energy-entropy framework by introducing the key quantities involved. Let $\pi$ be any portfolio strategy. Taking logarithm on both sides of \eqref{eqn:relativevalue} and using the notation $\Delta A(t) := A(t + 1) - A(t)$, we have
\begin{equation} \label{eqn:EEstep1}
\begin{split}
\Delta \log V_{\pi}(t) &= \log \left(\sum_{i = 1}^n  \pi_i(t) \frac{\mu_i(t + 1)}{\mu_i(t)}\right) \\
  &= \sum_{i = 1}^n \pi_i(t) \log \frac{\mu_i(t + 1)}{\mu_i(t)} + \\
  &\quad  \left( \log \left(\sum_{i = 1}^n  \pi_i \frac{\mu_i(t + 1)}{\mu_i(t)}\right) - \sum_{i = 1}^n \pi_i(t) \log \frac{\mu_i(t + 1)}{\mu_i(t)}\right).
\end{split}
\end{equation}

\begin{definition}[Discrete excess growth rate]
Let $\pi = \{\pi(t)\}_{t = 0}^{\infty}$ be a portfolio strategy. The discrete excess growth rate of $\pi$ for the period $[t, t + 1]$ is defined by
\begin{equation} \label{eqn:excessgrowth}
\gamma_{\pi}^*(t) = \log \left(\sum_{i = 1}^n  \pi_i(t) \frac{\mu_i(t + 1)}{\mu_i(t)}\right) - \sum_{i = 1}^n \pi_i(t) \log \frac{\mu_i(t + 1)}{\mu_i(t)}.
\end{equation}
The cumulative discrete excess growth rate is denoted by
\begin{equation} \label{eqn:cumexcessgrowth}
\Gamma_{\pi}^*(t) = \sum_{s = 0}^{t - 1} \gamma_{\pi}^*(s).
\end{equation}
\end{definition}

Note that $\gamma_{\pi}^*(t)$ depends on $\pi(t)$, $\mu(t)$ and $\mu(t + 1)$, and can be computed at time $t + 1$. By Jensen's inequality, the discrete excess growth rate $\gamma_{\pi}^*(t)$ is always non-negative. More explicitly, consider a random variable $Y$ such that $Y = \log \frac{\mu_i(t + 1)}{\mu_i(t)}$ with probability $\pi_i(t)$. Then we have
\[
\gamma_{\pi}^*(t) = \log {\Bbb E}_{\pi(t)} \left(e^{Y - {\Bbb E}_{\pi(t)} Y} \right) \geq 0,
\]
where ${\Bbb E}_{\pi(t)}$ denotes the expectation with respect to $\pi(t)$ viewed as a probability distribution. In particular, $\gamma_{\pi}^*(t)$ is strictly positive unless $Y$ is $\pi(t)$-a.s.~constant. This allows us to view $\gamma_{\pi}^*(t)$ as a measure of {\it cross-sectional volatility} of the market.

By Taylor approximation, we have
\begin{equation}
\gamma_{\pi}^*(t) \approx \frac{1}{2} \sum_{i, j = 1}^n \pi_i(t) \left(\delta_{ij} - \pi_j(t)\right) \Delta \log \mu_i(t) \Delta \log \mu_j(t)
\end{equation}
when $\mu(t + 1)$ is close to $\mu(t)$ (here $\delta_{ij}$ is the Kronecker delta). In the limit, this becomes the {\it excess growth rate} introduced in \cite{FS82} (see also \cite[(1.13)]{FKSurvey}). Since we work in discrete time, for simplicity we will drop the word `discrete' and this should cause no confusion with the continuous time excess growth rate.

The excess growth rate is {\it num\'{e}raire-invariant} in the following sense (see also \cite[Lemma 1.3.4]{F02}).

\begin{lemma} [Num\'{e}raire invariance] \label{lem:invariance}
Let $X_i(t)$ be the capitalization of stock $i$ as in \eqref{eqn:marketweight}, and let $\{M(t)\}_{t = 0}^{\infty}$ be any positive sequence serving as the num\'{e}raire. Let $\widetilde{r}_i(t) = \Delta \log (X_i(t) / M(t))$ be the log return of stock $i$ relative to $M(t)$, and $\widetilde{r}(t) = \Delta \log ( Z_{\pi}(t) / M(t) )$ be the corresponding quantity of the portfolio. Then
\begin{equation} \label{eqn:invariance}
\gamma_{\pi}^*(t) = \widetilde{r}(t) - \sum_{i = 1}^n \pi_i(t) \widetilde{r}_i(t).
\end{equation}
\end{lemma}
\begin{proof}
By \eqref{eqn:portfoliovalue}, we have
\[
\widetilde{r}(t) = \log\left( \sum_{i=1}^n \pi_i(t) \frac{X_i(t+1)}{X_i(t)}\cdot \frac{M(t)}{M(t+1)}  \right) =\Delta \log Z_{\pi}(t) - \Delta \log M(t).
\]
In the same way
\[
\begin{split}
\sum_{i=1}^n \pi_i(t) \widetilde{r}_i(t) &= \sum_{i=1}^n \pi_i(t) \log \frac{X_i(t+1)}{X_i(t)} - \Delta \log M(t)\\
&= \sum_{i = 1}^n \pi_i(t) \log \frac{\mu_i(t + 1)}{\mu_i(t)} + \Delta \log Z_{\mu}(t) - \Delta \log M(t).
\end{split}
\]
The lemma is proved by taking the difference of the two equations.
\end{proof}

In particular, taking $M(t) \equiv 1$ in \eqref{eqn:invariance}, we have
\[
\gamma_{\pi}^*(t) = \log\left(1 + \sum_{i = 1}^n \pi_i(t) R_i(t) \right) - \sum_{i = 1}^n \pi_i(t) \log \left(1 + R_i(t)\right),
\]
which the difference between the portfolio's logarithmic return and the weighted average of the stock's logarithmic returns. Thus $\gamma_{\pi}^*(t)$ is also equal to the {\it diversification return} studied in \cite{BF92}. 


We interpret the cumulative excess growth rate $\Gamma_{\pi}^*(t)$ as the amount of market cross-sectional volatility that can potentially be captured by the portfolio strategy $\pi$. Since $\gamma_{\pi}^*(t)$ is usually strictly positive (when $\pi(t) \in \Delta_n$) and the market weight $\mu(t)$ is constantly fluctuating, we expect that $\Gamma_{\pi}^*(t) \uparrow \infty$ as $t \uparrow \infty$. The term `energy' refers to this volatility term. More discussion about the terminology will be given in Section \ref{sec:generalEE}.

\subsection{Relative entropy}
The second key quantity of the energy-entropy framework is the {\it relative entropy} originating from information theory.

\begin{definition} [Relative entropy]
For $p \in \overline{\Delta}_n$ and $q \in \Delta_n$, the relative entropy $H\left( p \mid q\right)$ is defined by
\[
H\left( p \mid q\right) = \sum_{i = 1}^n p_i \log \frac{p_i}{q_i},
\]
with the convention $0 \log 0 = 0$.
\end{definition}

Relative entropy is also called the Kullback-Leibler divergence in statistics. Again by Jensen's inequality we have $H\left( p \mid q\right) \geq 0$, and $H\left( p \mid q\right) = 0$ if and only if $p = q$. Relative entropy can be interpreted as a kind of distance between probability vectors although it is not symmetric and does not satisfy the triangle inequality. For further properties of relative entropy we refer the reader to \cite[Chapter 2]{CT06}.

Now consider the first term of the last line of \eqref{eqn:EEstep1}. Observe that
\begin{equation} \label{eqn:EEstep2}
\begin{split}
\sum_{i = 1}^n \pi_i(t) \log \frac{\mu_i(t + 1)}{\mu_i(t)} &= \sum_{i = 1}^n \pi_i(t) \log \frac{\mu_i(t + 1)}{\pi_i(t)} - \sum_{i = 1}^n \pi_i(t) \log \frac{\pi_i(t)}{\mu_i(t)} \\
 &= - H\left( \pi(t) \mid \mu(t + 1) \right) + H\left( \pi(t) \mid \mu(t) \right).
\end{split}
\end{equation}
Combining \eqref{eqn:EEstep1} and \eqref{eqn:EEstep2}, we can decompose the relative log return $\Delta \log V_{\pi}(t)$ in the form
\begin{equation} \label{eqn:EEstep3}
\Delta \log V_{\pi}(t) = \gamma_{\pi}^*(t) + \left(H\left( \pi(t) \mid \mu(t + 1) \right) - H\left( \pi(t) \mid \mu(t) \right)\right).
\end{equation}

\subsection{Constant-weighted portfolios}
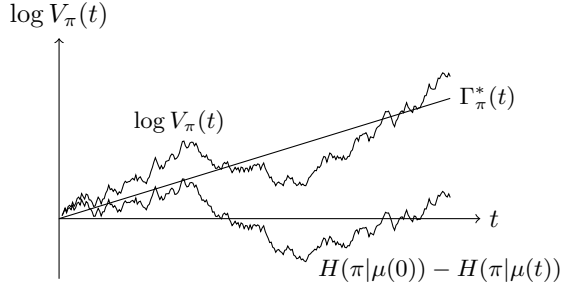
\begin{figure}[t!]
\centering
\begin{tikzpicture}[scale = 0.4]

\draw [<->] (0, 6) -- (0, 0) -- (14, 0);
\draw       (0, 0) -- (0, -2);
\node [right] at (14, 0) {$t$};
\node [above] at (0, 6) {$\log V_{\pi}(t)$};



\draw (0, 0) -- (13, 4);
\node [right] at (13, 4) {\small $\Gamma_{\pi}^*(t)$};
\node [above] at (4, 2.5) {\small $\log V_{\pi}(t)$};
\node [right] at (8.3, -1.7) {\small $H(\pi | \mu(0)) - H(\pi | \mu(t)) $};


\pgfmathsetseed{8}
\draw (0.1, 0.1)
        \foreach \i
        [
                evaluate = \W using rand - 0.5,
                evaluate = \x using 0.05,
                evaluate = \y using 0.2*\W + 0.107
        ]
        in {1,..., 258}{ -- ++(\x, \y)};
        
\pgfmathsetseed{8}
\draw (0.1, 0.1)
        \foreach \i
        [
                evaluate = \W using rand - 0.5,
                evaluate = \x using 0.05,
                evaluate = \y using 0.2*\W + 0.0914
        ]
        in {1,..., 258}{ -- ++(\x, \y)};
\end{tikzpicture}
\caption{Energy-entropy decomposition of a constant-weighted portfolio $\pi$.} \label{fig:EEcw}
\end{figure}

Constant-weighted portfolios are the simplest and conceptually most important in the theory of rebalancing. Let us first apply the decomposition \eqref{eqn:EEstep3} to a constant-weighted portfolio $\pi \in \overline{\Delta}_n$. Summing \eqref{eqn:EEstep3} over time, we have the {\it energy-entropy decomposition}
\begin{equation} \label{eqn:EEcw}
\log V_{\pi}(t) = \Gamma_{\pi}^*(t) + \left(H\left(\pi \mid \mu(0)\right) - H\left(\pi \mid \mu(t)\right)\right),
\end{equation}
where `energy' refers to the volatility term $\Gamma_{\pi}^*(t)$ (see Section \ref{sec:generalEE} for more discussion). Note that the sum over the relative entropies is a telescoping sum because $\pi(t) \equiv \pi$ is constant over time. This decomposition is illustrated in Figure \ref{fig:EEcw}.

\begin{remark}
For constant-weighted portfolios, the decomposition \eqref{eqn:EEcw} is a discrete time version of Fernholz's `master equation' (see \cite[Theorem 3.1.5]{F02}) for functionally generated portfolios. As a matter of fact, constant-weighted portfolios are functionally generated (where the `generating function' is the geometric mean). The general energy-entropy decomposition \eqref{eqn:EEstep4} generalizes the decomposition to any dynamic portfolio strategy. See \cite{PW14} for a discrete time, pathwise approach to the theory of functionally generated portfolios.
\end{remark}

In the decomposition \eqref{eqn:EEcw}, the cumulative excess growth rate $\Gamma_{\pi}^*(t)$ measures the amount of market volatility captured by the portfolio $\pi$ (which is analogous to the number of matched factors in the example in the Introduction). The relative entropy term
\[
H\left(\pi \mid \mu(0)\right) - H\left(\pi \mid \mu(t)\right)
\]
measures how much the relative performance deviates from $\Gamma_{\pi}^*(t)$. Note that the relative entropy term depends only on the initial and current positions of the market weight vector; it represents how the change in capital distribution affects the performance of the portfolio, excluding the effect of volatility. In particular, if the market weight vector becomes closer to the portfolio $\pi$ in the sense that $H\left( \pi \mid \mu(t) \right) < H\left( \pi \mid \mu(0) \right)$, the relative entropy term is positive and $V_{\pi}(t) > 1$.

In typical market situations, we expect that $\Gamma_{\pi}^*(t)$ grows linearly in time, i.e., $\Gamma_{\pi}^*(t) - \Gamma_{\pi}^*(s) \approx \epsilon(t - s)$ for some $\epsilon > 0$. While in the short run the fluctuation of $\log V_{\pi}(t)$ is dominated by the relative entropy term, long term growth comes from the cumulated excess growth rate. A simple example when this happens is given by the following proposition.

\begin{proposition} \label{prop:relativearbitrage1}
Let $\pi \in \overline{\Delta}_n$ be a constant-weighted portfolio. Suppose the market weight sequence $\{\mu(t)\}_{t = 0}^{\infty}$ satisfies the following conditions: (i) there exists a compact set $K \subset \Delta_n$ such that $\mu(t) \in K$ for all $t$, and (ii) $\Gamma_{\pi}^*(t) \uparrow \infty$ as $t \uparrow \infty$. Then $V_{\pi}(t) \uparrow \infty$ as $t \uparrow \infty$. In particular, let $t_0 = \inf\{t \geq 0: \Gamma_{\pi}^*(t) > c\}$ where $-c := \inf_{p \in K}  \left( H\left(\pi \mid \mu(0)\right) - H\left(\pi \mid p\right) \right)$. Then $V_{\pi}(t) > 1$ for all $t \geq t_0$.
\end{proposition}
\begin{proof}
Since $\mu(t) \in K$ for all $t$, by continuity of the relative entropy $H\left( \pi \mid \cdot \right)$ and compactness of $K$, we have, for all $t$,
\[
\inf_{t \geq 0} \left( H\left(\pi \mid \mu(0)\right) - H\left(\pi \mid \mu(t)\right) \right) \geq \inf_{p \in K}  \left( H\left(\pi \mid \mu(0)\right) - H\left(\pi \mid p\right) \right) = -c > -\infty.
\]
The rest follows immediately from the decomposition formula \eqref{eqn:EEcw}.
\end{proof}

The beauty of Proposition \ref{prop:relativearbitrage1} is that it is pathwise and completely free of stochastic modeling assumptions. To wit, long term outperformance follows whenever the sequence $\{\mu(t)\}_{t = 0}^{\infty}$ satisfies the path properties (i) and (ii). Proposition \ref{prop:relativearbitrage1} does not claim that a constant-weighted portfolio always beat the market portfolio. Rather, it gives an explicit set of sufficient conditions whose validity in a given market can be evaluated by the portfolio manager. Thus, the pathwise approach separates the problem of rebalancing into two parts: (i) the path properties required for rebalancing to be profitable, and (ii) whether these conditions are satisfied by the actual market over the investment horizon. Previous approaches tend to consider both questions together and give mixed results depending on the data used.

In an equity market there is no reason why $\mu(t)$ would stay within a compact subset of $\Delta_n$. In fact, in a typical market the market weights of most stocks are close to $0$. Proposition \ref{prop:relativearbitrage1} is more applicable, for example, if we interpret each asset as one or a group of industrial sectors or countries. In these situations we expect the capital distribution to be more stable.

As constant-weighted portfolios are of limited applicability, it is of interest to see how a dynamic portfolio strategy can be fitted in this framework. This is the purpose of the next subsection.

\subsection{General energy-entropy decomposition} \label{sec:generalEE}
For a general portfolio strategy $\pi$ where $\pi(t)$ may not be constant over time, we will rearrange \eqref{eqn:EEstep3} such that the effect of changing portfolio weights can be quanitifed. For this purpose, we write
\begin{equation} \label{eqn:EEstep4}
\begin{split}
\Delta \log V_{\pi}(t) &= \gamma_{\pi}^*(t) + \left(H\left( \pi(t) \mid \mu(t)\right) - H \left( \pi(t + 1) \mid \mu(t + 1) \right)\right) \\
&  \quad + \left(H\left( \pi(t + 1) \mid \mu(t + 1)\right) - H \left( \pi(t) \mid \mu(t + 1) \right)\right).
\end{split}
\end{equation}
The following nmemonic is helpful to remember and interpret the previous decomposition:
\begin{equation} \label{eqn:EEnames}
\Delta \log V_{\pi}(t) = \Delta \text{energy} - \Delta \text{relative entropy} + \Delta \text{control},
\end{equation}
where
\begin{equation} \label{eqn:EEterms}
\begin{split}
&\Delta \text{energy} = \gamma_{\pi}^*(t),\\
&\Delta \text{relative entropy} = H\left( \pi(t + 1) \mid \mu(t + 1) \right) - H \left( \pi(t) \mid \mu(t) \right), \\
&\Delta \text{control} = H\left( \pi(t + 1) \mid \mu(t + 1)\right) - H \left( \pi(t) \mid \mu(t + 1) \right).
\end{split}
\end{equation}
We call \eqref{eqn:EEstep4} and \eqref{eqn:EEnames} (as well as their time aggregates) the {\it energy-entropy decomposition} of the portfolio strategy $\pi$.

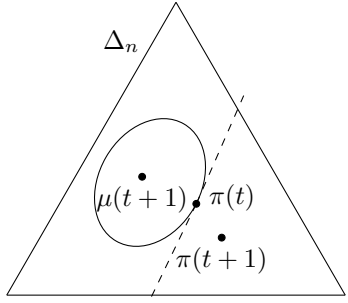
\begin{figure}[t!]
\begin{tikzpicture}[scale = 0.45]
\draw (0, 0) -- (10, 0);
\draw (0, 0) -- (5, 8.66);
\draw (5, 8.66) -- (10, 0);
\node [left] at (4.2, 7.5) {$\Delta_n$};

\draw [fill] (4, 3.5) circle [radius=0.1];
\node [below] at (4, 3.5) {$\mu(t+1)$};

\draw [name path=ellipse, rotate = -30]
                         (2, 5) ellipse (1.5cm and 2cm);
\draw [fill] (5.6, 2.7) circle [radius=0.1];

\node [right] at (5.6 + 0.1, 2.7 + 0.2) {$\pi(t)$};

\draw [dashed] (5.6, 2.7) -- (7, 5.9);
\draw [dashed] (5.6, 2.7) -- (4.2, -0.2);

\draw [fill] (5.6 + 0.75, 2.7 - 1) circle [radius=0.1];
\node [below] at (5.6 + 0.75, 2.7 - 1) {$\pi(t+1)$};
\end{tikzpicture}
\caption{Energy-entropy rebalancing. The oval shape is the contour of $H \left( \cdot \mid \mu(t + 1) \right)$, and the dotted line represents the tangent plane at $\pi(t)$. In the figure, $\pi(t + 1)$ is chosen such that $H \left( \pi(t + 1) \mid \mu(t + 1) \right) > H \left( \pi(t) \mid \mu(t + 1) \right)$. Thus the $\Delta \text{control}$ term is positive for this period.} \label{fig:EE}
\end{figure}

Some remarks of the terminologies are in order. Intuitively, we think that the market offers a constant amount of volatility in the form $\Gamma_{\pi}^*(t) \uparrow \infty$. Under suitable market conditions, it can be captured and turned into profit by a dynamic rebalancing strategy. This is analogous to energy in nature (such as sea waves and wind) that can be turned into work by machines. As the name suggests, the relative entropy term monitors the distance between the portfolio weight and the market weight vectors. Finally, the control term is determined by the new portfolio weight vector $\pi(t + 1)$ chosen at time $t + 1$. Since it depends on the action of the investor, we call it the control term. It is positive when $H \left( \pi(t + 1) \mid \mu(t + 1) \right) > H \left( \pi(t) \mid \mu(t + 1) \right)$, i.e., the portfolio moves away from the market (see Figure \ref{fig:EE}), and is negative when the portfolio moves towards the market. In the second case, we say that energy is `spent' to get closer to the market (see Example \ref{eg:market}).

\begin{example}[Constant-weighted portfolio]
Since $\pi(t) \equiv \pi$ for all $t$, for a constant-weighted portfolio the control term always satisfies $\Delta \text{control} \equiv 0$. Note that $\Delta \text{control} = 0$ does not mean that the portfolio does not trade at time $t + 1$ (see Remark \ref{re:rebalancing}). It simply means that the constant-weighted portfolio does not attempt to make $H\left( \pi(t + 1) \mid \mu(t + 1)\right)$ larger or small than $H\left( \pi(t) \mid \mu(t + 1)\right)$.
\end{example}

\begin{example}[Market portfolio] \label{eg:market}
The market portfolio satisfies $\pi(t) \equiv \mu(t)$ for all $t$. By definition, the relative entropy term is identically zero. Since $V_{\mu}(t) \equiv 1$ by definition, from \eqref{eqn:EEnames} we have
\[
\Delta \text{energy} = - \Delta \text{control}
\]
for all $t$. Continuing our metaphor about energy, we say that the market portfolio spends all the available energy to follow the market, and since no leftover energy (volatility) is accumulated, it can never outperform the market.
\end{example}

\section{Applications} \label{sec:applications}
\subsection{Energy-entropy portfolios} \label{sec:EEportfolios}
If the portfolio is not constant-weighted, we can choose the weights such that the $\Delta \text{control}$ term is either positive or negative. Write the time aggregate of the general energy-entropy decomposition \eqref{eqn:EEnames} in the form
\begin{equation} \label{eqn:EEregroup}
\begin{split}
\log V_{\pi}(t) &= \sum \left( \Delta \text{energy} + \Delta \text{control}\right) - \sum \Delta \text{relative entropy} \\
 &= \sum \left( \Delta \text{energy} + \Delta \text{control}\right) - \left( H \left( \pi(0) \mid \mu(0) \right) - H \left( \pi(t) \mid \mu(t)\right) \right).
\end{split}
\end{equation}
Intuitively, we think of $\Delta \text{energy} + \Delta \text{control}$ as the `leftover' of energy after rebalancing, and the purpose of such rebalancing is to control the relative entropy distance $H\left( \pi(t) \mid \mu(t) \right)$. This is because if $H\left( \pi(t) \mid \mu(t) \right)$ is large, a small change in market weights may cause a big drop in relative performance (this may be called {\it entropic risk} and is related to the concept of {\it tracking error}). The idea then is to choose a portfolio strategy such that $H\left( \pi(t) \mid \mu(t) \right)$ is bounded above and the first term is increasing. This leads to the following definition.

\begin{definition} [Energy-entropy portfolio]
An energy-entropy portfolio is a portfolio strategy $\pi$ which satisfies
\begin{equation} \label{eqn:EEportfolio}
\begin{split}
\Delta D(t - 1) &:= \Delta \text{energy}(t - 1) + \Delta \text{control}(t - 1)\\
   &= \gamma_{\pi}^*(t - 1) + \left( H \left(\pi(t) \mid \mu(t)\right) - H \left(\pi(t - 1) \mid \mu(t)\right)\right) \geq 0
\end{split}
\end{equation}
for all $t \geq 1$.
\end{definition}

Note that an energy-entropy portfolio can be constructed using only the observed history. At time $t$, the investor knows the previous portfolio weights $\{\pi(s) \}_{s = 0}^{t - 1}$ as well as the market weights $\{\mu(s)\}_{s = 0}^{t - 1}$ up to and including time $t$. From this, one can compute $\gamma_{\pi}^*(t - 1)$ and $H \left(\pi(t - 1) \mid \mu(t) \right)$, and then choose $\pi(t)$ for the period $[t, t + 1]$ such that $\Delta D(t - 1) \geq 0$.

Writing $D(0) = 0$, for an energy-entropy portfolio the decomposition \eqref{eqn:EEportfolio} allows us to write an expression similar to \eqref{eqn:EEcw} for constant-weighted portfolios:
\begin{equation} \label{eqn:EEdecomp5}
\log V_{\pi}(t) = D(t) + H \left(\pi(0) \mid \mu(0)\right) - H \left( \pi(t) \mid \mu(t)\right).
\end{equation}
We call $D(t)$ the {\it drift process}. As long as the drift process grows faster than the relative entropy distance $H \left(\pi(t) \mid \mu(t)\right)$ between the portfolio and the market weights, the energy-entropy portfolio eventually outperforms the market. 

\begin{remark}
In general, the portfolio weights of an energy-entropy portfolio depend on the entire history of market weights. For this reason, energy-entropy portfolios are typically not functionally generated (see \cite[Theorem 3.1.5]{F02} for the definition). Conversely, most functionally generated portfolios are not energy-entropy in the sense of \eqref{eqn:EEportfolio}. This is because a functionally generated portfolio $\pi$ is a deterministic function of the current market weight, i.e., $\pi(t) = \pi(\mu(t))$, so $\pi$ can be regarded as a portfolio map $\pi: \Delta_n \rightarrow \overline{\Delta}_n$. In fact, we proved in \cite{PW14} that among all deterministic portfolio maps, functionally generated portfolios are, in a certain sense, all the volatility capturing portfolios. Although the energy-entropy framework does not exhaust all possibilities in the dynamic case, it provides a systematic method of constructing volatility capturing portfolios.
\end{remark}

As an explicit example of energy-entropy portfolios, we introduce a family of portfolio strategies we call {\it $\lambda$-strategy}. The strategy depends on a parameter $\lambda \in [0, 1]$ and works as follows. Suppose we hold the portfolio $\pi(t)$ at time $t$. At time $t + 1$, we observe $\mu(t + 1)$ and the energy term $\gamma_{\pi}^*(t)$. We then move the portfolio towards $\mu(t + 1)$ so that $\pi(t + 1)$ is a convex combination of $\pi(t)$ and $\mu(t + 1)$, and the position is chosen so that we `consume' $\lambda$ fraction of the energy $\gamma_{\pi}^*(t)$, i.e., $\Delta \text{control} = -\lambda \Delta \text{energy}$. In other words, we are `saving' $1 - \lambda$ fraction of the energy term for each period. The idea is to rebalance towards the market in order to reduce the `entropic risk' (see Section \ref{sec:EEportfolios}), and the step size is bounded such that a constant proportion of market volatility is captured.

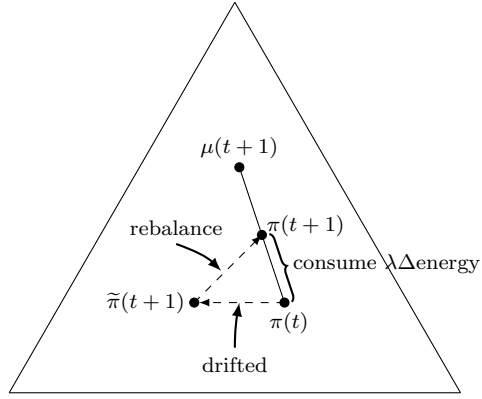
\begin{figure}[t!]
\centering
\begin{tikzpicture}[scale = 0.6]
\draw (0, 0) -- (10, 0);
\draw (0, 0) -- (5, 8.66);
\draw (5, 8.66) -- (10, 0);

\draw [fill] (6.1, 2) circle [radius = 0.1];
\node [below] at (6.2, 2) {\footnotesize $\pi(t)$};

\draw [fill] (5.1, 5) circle [radius = 0.1];
\node [above] at (5.1, 5) {\footnotesize $\mu(t+1)$};

\draw [fill] (4.1, 2) circle [radius = 0.1];
\node [left] at (4.1, 2) {\footnotesize $\widetilde{\pi}(t+1)$};
\draw [->, >=latex][dashed] (6.1, 2) -- (4.2, 2);
\node [below][black] at (5, 1) {\footnotesize drifted};
\draw[-latex,thick,black](5,1) to[out=90,in=260] (5.1,2);

\draw (6.1, 2) -- (5.1, 5);

\draw [fill] (5.6, 3.5) circle [radius = 0.1];
\draw [->, >=latex][dashed] (4.1, 2) -- (5.6, 3.5);
\node [above][black] at (3.7, 3.3) {\footnotesize rebalance};
\draw[-latex,thick,black](3.7, 3.3) to[out=340,in=135] (4.7, 2.7);
\node [right] at (5.5, 3.7) {\footnotesize $\pi(t+1)$};

\draw[decorate,thick,decoration={brace, raise=1pt, mirror}, xshift = 3pt][black]
     (6.1, 2) -- (5.6, 3.5) node [midway, xshift=2pt, yshift=2pt][right]{\footnotesize consume $\lambda \Delta \text{energy}$};

\end{tikzpicture}
\caption{Illustration of the $\lambda$-strategy. At time $t + 1$, the implied portfolio weights is $\widetilde{\pi}(t + 1)$ (see \eqref{eqn:implied}). The portfolio vector $\pi(t + 1)$ is a convex combination of $\pi(t)$ and $\mu(t + 1)$ chosen such that $\Delta \text{control} = -\lambda \Delta \text{energy}$.} \label{fig:lambda-strategy}
\end{figure}

Explicitly, we construct the portfolio by the following algorithm:
\begin{enumerate}
\item[(i)] Fix a starting weight vector $\pi(0) \in \overline{\Delta}_n$.
\item[(ii)] Suppose $\pi(t)$ has been chosen at time $t$. At time $t + 1$, $\mu(t + 1)$ is revealed and the discrete excess growth rate $\gamma_{\pi}^*(t)$ can be computed. Then define
\[
\pi(t + 1) = \pi(t) + \eta \left( \mu(t + 1) - \mu(s) \right)
\]
where $\eta$ solves the equation $ -\lambda \Delta \text{energy} = \Delta \text{control}$:
\begin{equation} \label{eqn:lambda}
- \lambda \gamma_{\pi}^*(t) = H \left( \pi(t + 1) \mid \mu(t + 1) - H \left( \pi(t) \mid \mu(t + 1) \right) \right).
\end{equation}
\end{enumerate}

\begin{example}
If $\lambda = 0$, then $\pi(t + 1) = \pi(0)$ for all $t$ and so $\pi$ is a constant-weighted portfolio. If $\lambda = 1$ and $\pi(0) = \mu(0)$, then $\pi$ is the market portfolio $\mu$.
\end{example}

By \eqref{eqn:lambda}, we immediately obtain the following energy-entropy decomposition of the $\lambda$-strategy.

\begin{proposition}
Let $\pi$ be the $\lambda$-strategy with $\lambda \in [0, 1]$. Then
\begin{equation} \label{eqn:EElambda}
\log V_{\pi}(t) = \left(H \left( \pi(0) \mid \mu(0)\right) - H \left( \pi(t) \mid \mu(t) \right)\right) + (1 - \lambda) \Gamma^*_{\pi}(t).
\end{equation}
\end{proposition}

The equation \eqref{eqn:lambda} is nonlinear. In practice, we can estimate the solution by linear approximation, and a more sophisticated version is to let $\lambda$ depend on market conditions. The actual performance of the $\lambda$-strategy will be studied in Section \ref{sec:empirical}.

\begin{remark}
The $\lambda$-strategy is analogous to the {\it diversity-weighted portfolio} given by \eqref{eqn:diversity}. Note then the diversity-weighted portfolio is equal-weighted when $\lambda = 0$ and is the market portfolio when $\lambda = 1$. For $\lambda \in [0, 1]$, the relative log return of the diversity-weighted portfolio can be expressed approximately as (see \cite[Example 3.4.4]{F02})
\begin{equation}
\Delta \log V_{\pi}(t) \approx \Delta \log \Phi(t) + (1 - \lambda) \gamma_{\pi}^*(t),
\end{equation}
where
\[
\Phi(t) = \left( \sum_{i = 1}^n \mu_i^{\lambda}(t) \right)^{1/\lambda}
\]
is the {\it diversity function}. Both families of portfolio can be regarded as an interpolation between a constant-weighted portfolio and a market-following portfolio.
\end{remark}

\subsection{Empirical examples} \label{sec:empirical}
In this subsection we illustrate the $\lambda$-strategy defined in Section \ref{sec:EEportfolios} using actual data.

\begin{figure}[t!]  
\centering
\includegraphics[scale=0.5]{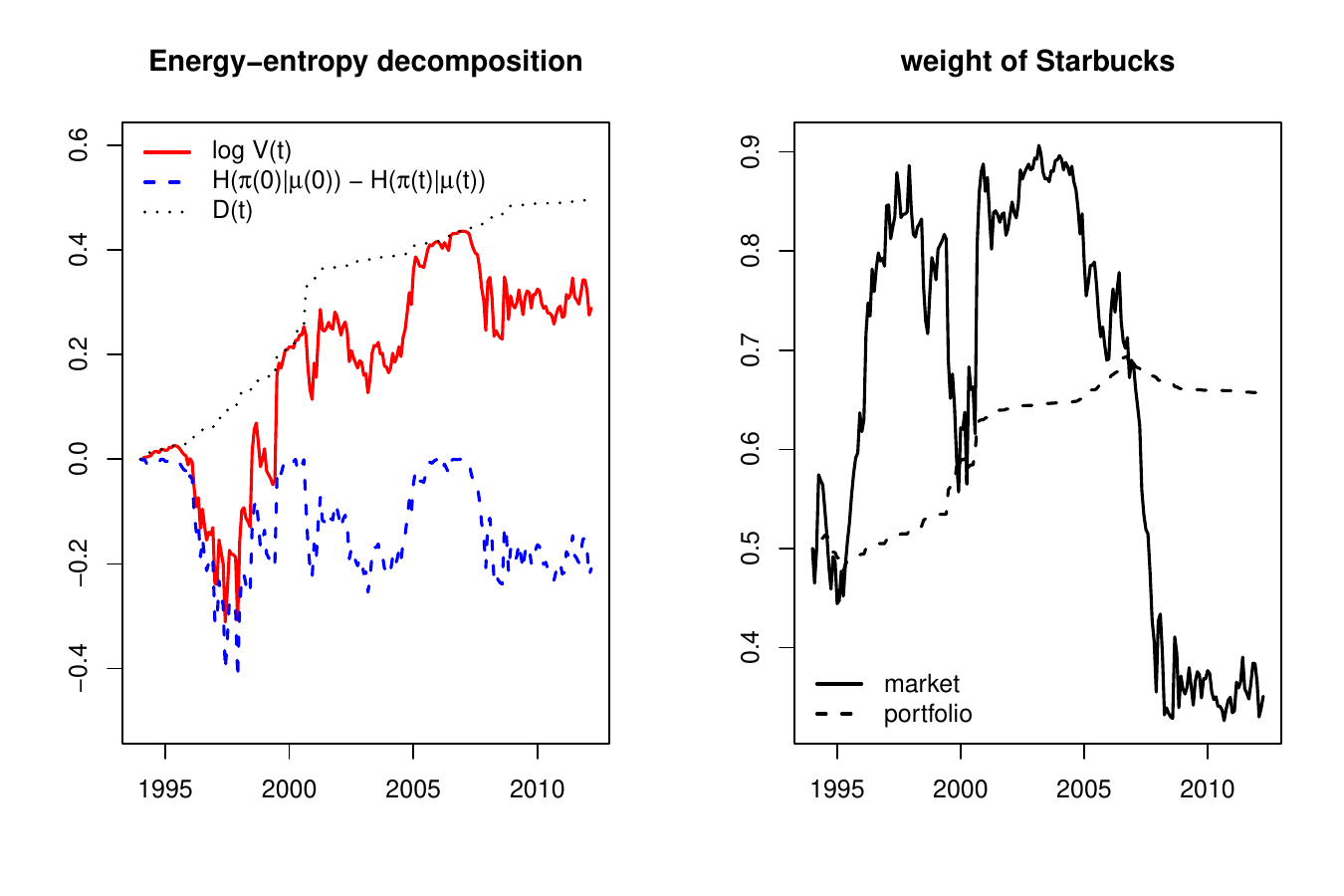}
\vspace{-0.5cm}
\caption{(Left) Energy-entropy decomposition of $\pi$ with respect to the Apple-Starbucks market, where $\lambda = 0.3$. (Right) Portfolio and market weights of Starbucks as a function of time.}
\label{fig:applestarbucks}
\end{figure}

As a first example we consider the monthly stock prices of Apple and Starbucks from January 1994 to April 2012. The market consists of these two stocks and we normalize the prices so that they are equally weighted in January 1994 (i.e., $\mu(0) = (0.5, 0.5)$). This data set is also studied in \cite{PPA12}. Now we simulate the performance of the $\lambda$-strategy with $\lambda = 0.3$. We let $\pi(0) = \mu(0) = (0.5, 0.5)$ be the starting weights and use monthly time steps. Figure \ref{fig:applestarbucks} (left) plots the energy-entropy decomposition
\[
\log V_{\pi}(t) = D(t) + H\left(\pi(0) \mid \mu(0)\right) - H \left( \pi(t)\mid \mu(t)\right)
\]
as a function of time. Since $\pi$ is an energy-entropy portfolio, the drift process $D(t)$ is increasing by construction. From the figure, it is clear that the drift process drives the long term outperformance of the portfolio. On the right we also plot the weight of Starbucks. We see that the portfolio moves towards the market slowly (approximating a finite variation process); it adjusts more rapidly when the market is volatile, i.e., when the energy term is large. Compared to a constant-weighted portfolio, here the fluctuation of the relative entropy term is smaller.

\begin{figure}[t!]
\centering
\includegraphics[scale=0.55]{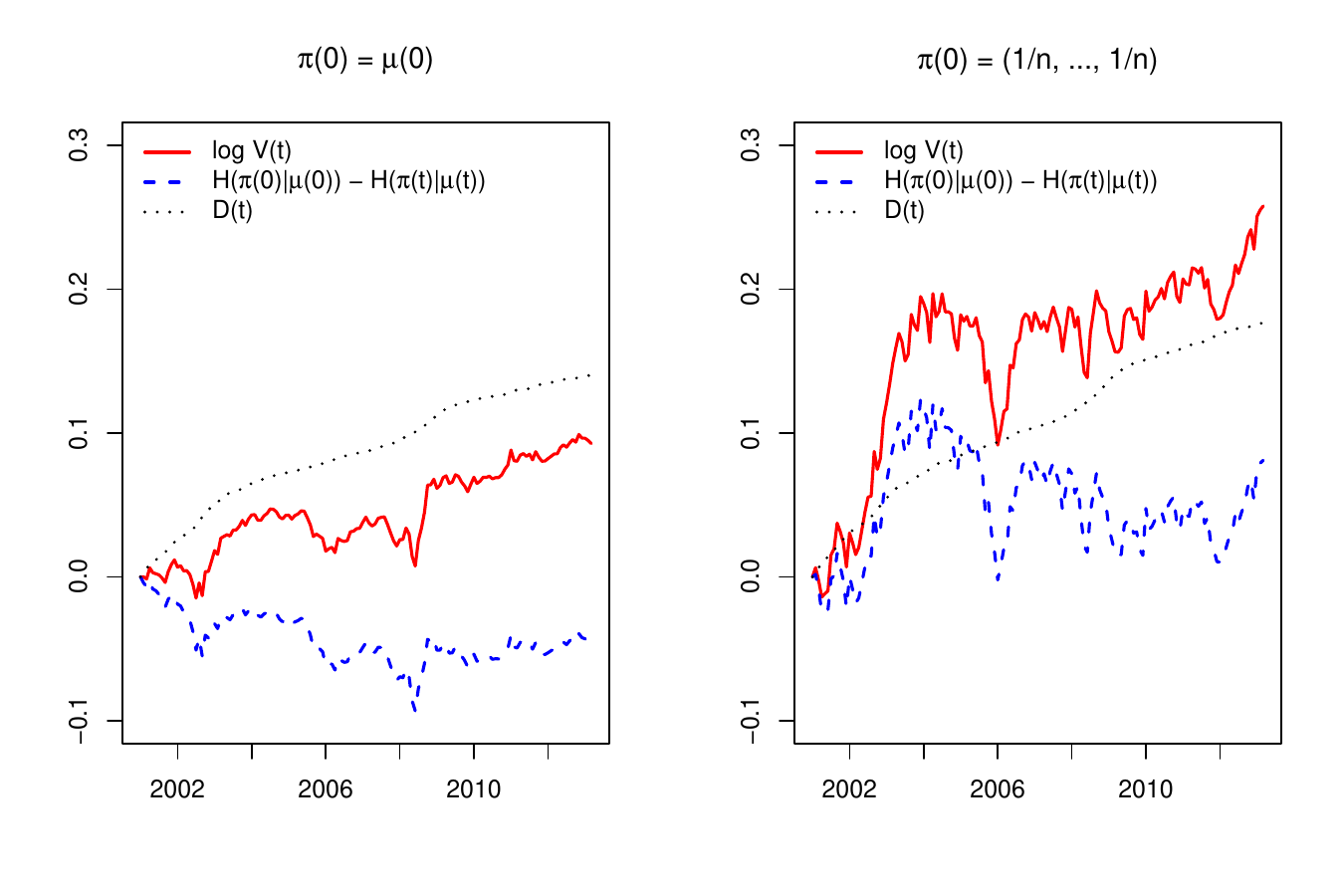}
\caption{Performance of $\pi$ relative to the hypothetical market of emerging countries, again with $\lambda = 0.3$. (Left) The portfolio begins at the market weights. (Right) The portfolio is equal weighted initially.}
\label{fig:emergingmarkets}
\end{figure}

Next we turn to a more realistic example where we consider monthly country returns (in US dollars) of 18 emerging market countries from January 2001 to March 2013. More precisely, for each country we pick a country index, and the returns of that index are taken to be the country returns. The data is extracted from Factset. The market consists of these countries, where the starting market weights are proportional to the total capitalizations of the indices. For example, the beginning market weights of Brazil, Chile and China are respectively 0.138, 0.044 and 0.073.

Again we simulate the performance of the $\lambda$-strategy with $\lambda = 0.3$. We consider two cases where $\pi(0) = \mu(0)$ and $\pi(0) = (1/n, \ldots, 1/n)$ respectively. Figure \ref{fig:emergingmarkets} plots the energy-entropy decompositions of the two portfolios. In both cases $\pi$ outperforms the market and the drift process has a steady increasing trend. Note that although we use the same update rule in both cases, the time series of the relative entropy terms are quite different because the portfolio depends on the entire history of portfolio and market weights.

\subsection{Hierarchical portfolios}\label{sec:hierarchical} 
Consider an investor in global markets. It is convenient to think of the portfolio in a hierarchical framework: (i) a portfolio that describes the proportion of capital invested in each country, (ii) for each country with several sectors in its economy, a portfolio that describes the amount invested in each sector as a proportion of the total money invested in that country, and finally (iii), for each sector of every country, how the allocated amount is distributed among various stocks as proportions of the corresponding total. The above is an example of a {\it hierarchical portfolio}, and similar structures arise for example in managing a fund of funds or combining performances of different managers. 

The portfolio weights of a hierarchical portfolio can be thought of naturally as conditional probabilities. In this subsection, we show that the discrete excess growth rate and relative entropy satisfy {\it chain rules} which allow us to quantify the profit or loss at each level of the hierarchy. For convenience, we study a two-step hierarchy (`sectors' and `stocks') and similar considerations can be applied to multiple levels.

Suppose there are $m$ sectors and each sector has $n_i$ stocks, $i = 1, \ldots, m$. The universe thus consists of $n \leq n_1 + \cdots + n_m$ stocks, with equality when the sectors are disjoint. Now a portfolio vector $\pi$ in this universe  is a combination of the {\it sector portfolios} $\pi_i = (\pi_{i1}, \ldots, \pi_{in_i}) \in \overline{\Delta}_{n_i}$. It is helpful notationally to regard each $\pi_i$ as a portfolio weight vector in $\Delta_n$ simply by putting zeros for stocks not in the sector. Let the sector weights be $\lambda = (\lambda_1, \ldots, \lambda_m) \in \overline{\Delta}_m$. Then the portfolio weight vector $\pi$ can be expressed in the form
\[
\pi = \sum_{i = 1}^n \lambda_i \pi_i.
\]

\begin{proposition}[Chain rules] \label{prop:chain}
Consider two pairs of sector weights and sector portfolios $\left(\lambda_i, \pi_i\right)$, $i = 1, \ldots, m$, and $\left(\alpha_i, \nu_i \right)$, $i = 1, \ldots, m$. Let $\pi$ and $\nu$ be the total portfolios as distributed over all $n$ stocks. We have the following identities.
\begin{enumerate}
\item[(i)] Chain rule for relative entropy:
\[
H \left( \pi \mid \nu \right) = H \left( \lambda \mid \alpha \right) + \sum_{i = 1}^m \lambda_i H \left( \pi_i \mid \nu_i \right).
\]
\item[(ii)] Chain rule for discrete excess growth rate:
\[
\gamma_{\pi}^*(t) = \gamma_{\pi}^{*, \text{sector}}(t) + \sum_{i = 1}^n \lambda_i(t) \gamma_{\pi_i}^{*, \text{stock}}(t).
\]
\end{enumerate}
Here, $\gamma_{\pi}^*(t)$ and $\gamma_{\pi}^{*, \text{stock}}$ are the discrete excess growth rate where the basic assets are the stocks, and $\gamma_{\pi}^{*, \text{sector}}$ is the discrete excess growth rate where the basic assets are the sector portfolios.
\end{proposition}
\begin{proof}
(i) The chain rule for relative entropy is a well-known property of relative entropy and can be found, for example, in \cite[page 24]{CT06}.

(ii) Fix a time period $[t, t + 1]$ and for notational simplicity we will drop time in the following computation. If $X(t)$ represents the market capitalization of a stock, its logarithmic return over the interval is $r(t) = \Delta \log X(t)$. Now let $r^{\text{stock}}$ be the vector of the logarithmic returns of all stocks. Similarly, let $r^{\text{sector}} = \left(r^{\text{sector}}_1, \ldots, r^{\text{sector}}_m\right)$ be the vector of sector logarithmic returns. We now use Lemma \ref{lem:invariance} repeatedly.

If we think of $\pi$ as a portfolio of the $n$ stocks, we can write
\begin{eqnarray} \label{Two stage Eqn 1}
r_{\pi} = \sum_{i=1}^m \sum_{j=1}^{n_i} \lambda_i \pi_{ij}  r_{ij}^{\text{stock}} + \gamma_{\pi}^{*}.
\end{eqnarray}
We may also think of $\pi$ as a mixture of sector portfolios, and each sector portfolio $\pi_i$ is a mixture of sector stocks. Hence, at the sector level we may also write
\begin{eqnarray} \label{Two stage Eqn 2}
r_{\pi} = \sum_{i=1}^m \lambda_i  r_i^{\text{sector}} + \gamma_{\pi}^{*, \text{sector}}
\end{eqnarray}

Finally, for each sector portfolio, we have
\begin{eqnarray} \label{Two stage Eqn 3}
r^{\text{sector}}_i = r_{\pi_i} = \sum_{j=1}^{n_i} \pi_{ij} r_{ij}^{\text{stock}} + \gamma_{\pi_i}^{*, \text{stock}}
\end{eqnarray}

Let $a\cdot b$ denote the Euclidean inner product of two vectors $a$ and $b$. Putting (\ref{Two stage Eqn 3}) into (\ref{Two stage Eqn 2}), we get
\begin{equation} \label{Two stage Eqn 4}
\begin{split}
r_{\pi} &= \sum_{i = 1}^m \lambda_i (\pi_i \cdot r^{\text{stock}} + \gamma_{\pi_i}^{*, \text{stock}}) +
                 \gamma_{\pi}^{*, \text{sector}} \\
             &= \pi \cdot r^{\text{stock}} + \sum_{i = 1}^m \lambda_i \gamma_{\pi_i}^{*, \text{stock}} + \gamma_{\pi}^{*, \text{sector}}.
\end{split}
\end{equation}

Comparing \eqref{Two stage Eqn 4} and \eqref{Two stage Eqn 1}, we have $\gamma_{\pi}^{*} = \sum_{i = 1}^m \lambda_i \gamma_{\pi_i}^{*, \text{stock}} + \gamma_{\pi}^{*, \text{sector}}$.
\end{proof}

Proposition \ref{prop:chain} provides a neat way of attributing the energy and relative entropy terms of any portfolio across levels. We now ask a natural question: if we run an energy-entropy portfolio within each sector and run an energy-entropy portfolio among the sectors, is the total portfolio also an energy-entropy portfolio? The following proposition gives a set of sufficient conditions when this is the case.

\begin{proposition}\label{thm:attribution}
Using the notation of Lemma \ref{prop:chain}, let $\pi(t) = \sum_{i=1}^m \lambda_i(t) \pi_i(t)$ be the portfolio and let $\mu(t) = \sum_{i=1}^m \alpha_i(t) \mu_i(t)$ denote the market portfolio (here each $\mu_i$ represents a sector market portfolio). Suppose each $\pi_i$ is an energy-entropy portfolio within sector $i$. Then $\pi$ is an energy-entropy portfolio over the entire universe if any of the following two conditions is satisfied.
\begin{enumerate}
\item[(i)] $\lambda$ is a constant-weighted portfolio.
\item[(ii)] $\lambda$ is an energy-entropy portfolio that satisfies the following monotonicity condition for any pair of indices $(i, j)$:
\begin{equation}\label{eq:monotonecond}
\begin{split}
\frac{\lambda_i(t+1)}{\lambda_i(t)} &\ge \frac{\lambda_j(t+1)}{\lambda_j(t)},\quad \text{if} \\
H \left(\pi_i(t+1)\mid \mu_i(t+1)\right) &> H \left(\pi_j(t+1) \mid \mu_j(t+1)\right).
\end{split}
\end{equation}
\end{enumerate}
\end{proposition}

\begin{proof}
We start at the decomposition \eqref{eqn:EEdecomp5}. It suffices to consider the drift process $D(t)$ and show that it is increasing in time. We have
\eq\label{eq:att1}
\begin{split}
\Delta D(t) &= \gamma_\pi^*(t) + H \left(\pi(t+1) \mid \mu(t+1)\right) - H \left( \pi(t) \mid \mu(t+1)\right).
\end{split}
\en

We now use Lemma \ref{prop:chain} to expand each term on the right side of the above equation:
\begin{eqnarray*}
\gamma_\pi^*(t) &=& \gamma^{*,\text{sector}}_\pi(t) + \sum_{i=1}^m \lambda_i(t) \gamma^{*,\text{stock}}_{\pi_i}(t),\\
H \left( \pi(t+1) \mid \mu(t+1)\right) &=& H \left(\lambda(t+1) \mid \alpha(t+1)\right) + \sum_{i=1}^m \lambda_i(t+1) H\left(\pi_i(t+1) \mid \mu_i(t+1)\right),\\
H \left( \pi(t) \mid \mu(t+1) \right) &=& H \left( \lambda(t) \mid \alpha(t+1)\right) + \sum_{i=1}^m \lambda_i(t) H \left(\pi_i(t) \mid \mu_i(t+1)\right).
\end{eqnarray*}

When $\lambda$ is constant-weighted, we have $\lambda(t)=\lambda(t+1)\equiv \lambda$. This allows us to combine the three terms above and get
\[
\begin{split}
\Delta D(t) &= \Delta D^{\text{sector}}(t) + \sum_{i=1}^m \lambda_i \Delta D^{\text{stock}}_i(t),
\end{split}
\]
where the notation is self-explanatory. Since we have energy-entropy portfolios within each sector and a constant weighted portfolio among the sectors, each of the $\Delta D^{\cdot}_{\cdot}(t)$ terms is non-negative. This shows that $\Delta D(t) \ge 0$ and proves that $\pi$ is an entropic-rebalancing portfolio.

In the general case we write
\[
\begin{split}
\Delta D(t) &= \Delta D^{\text{sector}}(t) + \sum_{i=1}^m \lambda_i(t+1) H \left(\pi_i(t+1) \mid \mu_i(t+1)\right) \\
&+ \sum_{i=1}^m \lambda_i(t)\left[  \gamma^{*,\text{stock}}_{\pi_i}(t) - H \left( \pi_i(t) \mid \mu_i(t+1)\right) \right]\\
&= \Delta D^{\text{sector}}(t) + \sum_{i=1}^m\left[  \lambda_i(t+1) - \lambda_i(t) \right] H \left( \pi_i(t+1) \mid \mu_i(t+1)\right)\\
& + \sum_{i=1}^m \lambda_i(t)\left[ H \left(\pi_i(t+1) \mid \mu_i(t+1)\right) + \gamma^{*,\text{stock}}_{\pi_i}(t) - H \left(\pi_i(t) \mid \mu_i(t+1)\right) \right].
\end{split}
\]

Because we run energy-entropy portfolios within and among the sectors, the first and the third term in the final expression above is nonnegative. We now show how to control the middle term.

Consider a random integer $I$ that takes value $i$ with probability $\lambda_i(t)$, for $i=1,2,\ldots,m$. Consider two functions:
\[
f(i) = \frac{\lambda_i(t+1)}{\lambda_i(t)}, \quad g(i)=  H \left(\pi_i(t+1) \mid \mu_i(t+1)\right).
\]
Obviously $E f(I)=1$ and, hence
\[
 \sum_{i=1}^m\left[  \lambda_i(t+1) - \lambda_i(t) \right] H \left(\pi_i(t+1) \mid \mu_i(t+1) \right)= \text{Cov}\left( f(I), g(I) \right).
\]

If $(I,I')$ are i.i.d.~random variables, the above covariance is given by the symmetrized expression
\[
 \text{Cov}\left( f(I), g(I) \right)= \frac{1}{2} E\Big[\left( f(I) - f\left(I'\right)  \right)\left( g(I) - g\left(I'\right) \right)\Big].
\]
Thus, under the assumed monotonicity condition \eqref{eq:monotonecond}, for any pair of values of $(I,I')$ we must have
\[
\left( f(I) - f\left(I'\right)  \right)\left( g(I) - g\left(I'\right) \right) \ge 0.
\]
This, in turn, implies that $ \text{Cov}\left( f(I), g(I) \right)\ge 0$. Combining everything we get $\Delta D(t) \ge 0$ and hence our result is proved.
\end{proof}

\appendix
\section{Rebalancing and functionally generated portfolios} \label{sec:fgp}
We continue the discussion in Section \ref{sec:binary}.

\begin{example}[Deterministic portfolio function]
Consider a portfolio strategy $\pi(t) = \left(q(Y(t)), 1 - q(Y(t))\right)$, where $q: {\Bbb R} \rightarrow [0, 1]$ is a function of $Y(t)$. Now the gain or loss from a match depends on the value of $Y$. The contribution of a matching from an up move from $k\sigma$ to $(k + 1)\sigma$ and a down move from $(k + 1)\sigma$ to $k\sigma$ is
\[
\left(1 + q(k\sigma)\left(e^{\sigma} + 1\right)\right) \left(1 + q((k + 1)\sigma)\left(e^{-\sigma} + 1\right)\right).
\]
When is this contribution greater than $1$? Assuming $q$ is differentiable and using Taylor approximation, if we let $\sigma \rightarrow 0$ we get the differential inequality
\begin{equation} \label{eqn:diffinequality}
q'(y) \leq q(y)(1 - q(y)).
\end{equation}
Note that if $q(y) = \frac{e^{y}}{1 + e^{y}}$ (i.e., when $\pi$ is the market portfolio), then $q' = q(1 - q)$ and equality holds. As it turns out, the inequality \eqref{eqn:diffinequality} is closely related to Fernholz's functionally generated portfolio. Using the definition of functionally generated portfolio given in \cite[Chapter 3]{F02}, we have the following result.
\end{example}

\begin{proposition} \label{eqn:FGP}
Any portfolio $\pi(t) = (q(Y(t)), 1 - q(Y(t)))$, where $q$ is continuously differentiable, is functionally generated. A
generating function (unique up to a multiplicative constant) is
\begin{equation} \label{eqn:genfunction}
\Phi(\mu_1, \mu_2) = \exp\left(F\left(\log \frac{\mu_1}{\mu_2} \right)
- \log \frac{1}{\mu_2} \right),
\end{equation}
where $F$ is an antiderivative of $q$. Moreover, the inequality \eqref{eqn:diffinequality} holds for all $y$ if and only if $\Phi$ is concave.
\end{proposition}
\begin{proof}
Let $\Phi = e^G$, where $G$ is a differentiable function of $y = \log \frac{X_1}{X_2} = \log \frac{\mu_1}{\mu_2}$. By \cite[(11.1)]{FKSurvey}, $\Phi$ generates the portfolio $(\pi_1, \pi_2)$ where
\[
\frac{\pi_1(\mu)}{\mu_1} = D_1 \log \Phi(\mu) + 1 - \mu_1(\mu) D_1 \log \Phi - \mu_2 D_2
\log \Phi = \frac{G'(y) + \mu_1}{\mu_1}.
\]
To generate the weights $(q, 1 - q)$, we require that
\[
q(y) = G'(y) + \mu_1 = G'(y) + \frac{e^y}{1 + e^y}.
\]
We may then pick $G(y) = F(y) - \int \frac{e^y}{1 + e^y} dy = F(y) - \log(1 + e^y)$. The second statement then follows by direct differentiation.
\end{proof}

The inequality \eqref{eqn:diffinequality} and Proposition \ref{eqn:FGP} are the beginning of a bigger story. The underlying `geometry of rebalancing' is studied in \cite{PW14}, giving rise to an elegant connection between functionally generated portfolio, convex analysis and optimal transport. Further development related to optimization and Cover's universal portfolio \cite{C91} are reported in \cite{W14} and \cite{W15}.

\section*{Acknowledgment}
This research was partially sponsored by Parametric Portfolio Associates at Seattle. We thank Paul Bouchey, Alex Paulsen, Mahesh Pritamani, Vassilii Nemtchinov, and David Stein for their comments, problems, and stimulating conversations. We are indebted to Robert Fernholz and Ioannis Karatzas for their detailed comments on an earlier draft. The first author would also like to thank Rik Sen from HKUST for many happy discussions. His research is also supported partially by NSF grant DMS 1308340.

\bibliographystyle{amsalpha}
\bibliography{infogeo}

\end{document}